\documentclass[%
 reprint,
 amsmath,amssymb,
 aps,
nofootinbib]{revtex4-1}

\usepackage{graphicx}
\usepackage{dcolumn}
\usepackage{bm}
\usepackage[symbol]{footmisc}

\usepackage{amsmath}
\usepackage{xcolor}
\usepackage{float}
\usepackage{multirow}
\usepackage{chemformula}

\usepackage{natbib}
\begin{document}

\preprint{APS/123-QED}

\title{\emph{Hop-Decorate}: An Automated Atomistic Workflow for Generating Defect Transport Data in Chemically Complex Materials}

\author{Peter Hatton}
\affiliation{Material Science and Technology Division, Los Alamos National Lab, Los Alamos, 87545, NM, USA.}

\author{Blas Pedro Uberuaga}
\affiliation{Material Science and Technology Division, Los Alamos National Lab, Los Alamos, 87545, NM, USA}

\author{Danny Perez}
\affiliation{Theoretical Division, Los Alamos National Lab, Los Alamos, 87545, NM, USA}

\date{\today}

\begin{abstract}
Chemically complex materials (CCMs) exhibit extraordinary functional properties but pose significant challenges for atomistic modeling due to their vast configurational heterogeneity. We introduce Hop-Decorate (HopDec), a high-throughput, Python-based atomistic workflow that automates the generation of defect transport data in CCMs. HopDec integrates accelerated molecular dynamics with a novel redecoration algorithm to efficiently sample migration pathways across chemically diverse local environments. The method constructs a defect-state graph in which transitions are associated with distributions of kinetic and thermodynamic parameters, enabling direct input into kinetic Monte Carlo and other mesoscale models. We demonstrate HopDec’s capabilities through applications to a Cu-Ni alloy and the spinel oxide (Fe,Ni)Cr$_2$O$_4$, revealing simple predictive relationships in the former and complex migration behaviors driven by cation disorder in the latter. These results highlight HopDec’s ability to extract physically meaningful trends and support reduced-order or machine-learned models of defect kinetics, bridging atomic-scale simulations and mesoscale predictions in complex material systems.
\end{abstract}

\maketitle

\section{Introduction}

The study of chemically complex materials (CCMs) has emerged as a pivotal theme across a diverse range of research fields, including structural materials development, energy conversion and storage, and nuclear technology \cite{Walck2024,Miracle2017,Zhang2014, Yang2022,Xiao2022}. These materials, characterized by their multi-principal element compositions and intricate chemical heterogeneity, exhibit unique mechanical, thermal, and electronic properties that are often unattainable in simpler, chemically homogeneous systems \cite{George2019,Gludovatz2014,Yeh2004}. A central aspect of understanding the performance and stability of CCMs is the characterization of point defect behavior, particularly their ability to accommodate and transport defects under operational conditions \cite{Granberg2016,Zhang2019a,Sickafus1995}.
Indeed, defect transport governs critical material properties such as diffusion, mechanical strength, radiation tolerance, and corrosion resistance \cite{Nastasi1996,Hatton2024}. In chemically simple materials, such as pure metals or stoichiometric oxides, the migration of defects can typically be described by a limited set of symmetrically unique stable states and transition pathways. Atomistic simulation methods-most notably molecular dynamics (MD) and transition state search algorithms-have been highly effective in these settings, providing insights into energy landscapes, diffusion constants, and activation barriers based on a relatively small number of migration mechanisms \cite{Voter2007,Henkelman2000,Thompson2022, Ebmeyer2024}.

In stark contrast to simple systems, CCMs present a formidable challenge for such modeling approaches. The lack of long-range chemical order in these materials results in a broad and continuous distribution of local atomic environments around defects. Even modest chemical complexity—for example, an equiatomic binary alloy—can give rise to thousands of distinct nearest-neighbor (NN) configurations, each potentially modifying the defect’s thermodynamics and kinetics \cite{Yang2023a,Xi2022,Xi2024}. This combinatorial explosion renders traditional exhaustive sampling methods impractical. Furthermore, environmental influences beyond the first coordination shell can significantly perturb migration pathways, demanding an even more extensive and nuanced exploration of the defect landscape.

In the study of defect migration, two key energetic metrics are particularly important: the energy barrier (E$_\text{b}$) and the energy difference between states ($\Delta$E). E$_\text{b}$ represents the kinetic hurdle that must be overcome for a defect to move between two configurations; it fundamentally governs the rate at which transitions occur via thermally activated processes. $\Delta$E, on the other hand, reflects the thermodynamic driving force favoring one defect configuration over another. Together, these quantities provide a complete energetic picture of both the kinetics and thermodynamics of defect behavior, making them essential targets for accurate modeling and predictive simulations.

Existing atomistic approaches face two major limitations when applied to CCMs. First, conventional MD simulations are often inefficient at sampling rare but critical migration events, especially at lower temperatures. Second, catalogue based methods, such as traditional Kinetic Monte Carlo techniques \cite{Lelievre2020,Kratzer} rely on pre-identifying a manageable number of migration mechanisms, an assumption that breaks down in the face of the massive configurational diversity present in CCMs. As a result, defect transport in these materials remains comparatively poorly understood and difficult to predict, hampering the development of predictive models for their long-term behavior \cite{Uberuaga2015,Voter2002}.

To address these challenges, a methodology is needed that can (i) efficiently generate a broad and representative sampling of defect migration pathways across the vast configurational space of CCMs, and (ii) encode this information in a way that is tractable for subsequent modeling, analysis, and integration into higher-scale simulations.

In this work, we introduce Hop-Decorate (HopDec), a Python-based, high-throughput atomistic workflow designed specifically for this purpose. HopDec systematically explores the defect state-space by combining accelerated MD techniques for stable state discovery with a novel \lq redecoration' strategy that samples local chemical environments. By leveraging embarrassingly parallel calculations and scalable data structures, HopDec maps the energy barriers and the thermodynamic stabilities of the associated states across diverse configurations with unprecedented efficiency.

Importantly, HopDec does not merely catalog transitions; it organizes them into an abstract defect-state graph where each transition is associated with a distribution of rates rather than a single value. This abstraction enables the direct construction of kinetic models, such as Kinetic Monte Carlo (KMC) simulations, that can faithfully capture the true variability of defect transport in chemically complex systems.

We demonstrate the power and flexibility of HopDec through two representative case studies. The first focuses on the Cu-Ni binary alloy system, a model chemically disordered FCC alloy where defect migration is known to be highly sensitive to local composition. Here, we uncover simple, physically interpretable relationships between local chemical environments and migration energetics, leading to predictive models with remarkable accuracy. The second case study applies HopDec to the spinel-structured complex oxide (Fe,Ni)Cr$_2$O$_4$, highlighting how HopDec accommodates systems with multiple, chemically distinct sublattices and varying cation orderings. These examples illustrate not only HopDec’s broad applicability but also its ability to generate new physical insights into defect transport mechanisms.

\section{Methodology}

\subsection{Hop-Decorate (HopDec)}

\textit{Hop-Decorate }(\textit{HopDec}) is a Python-based, high-throughput atomistic simulator designed to approximate the defect state-space in chemically complex materials. It generates a blueprint of the state-space based on a static realization of a given material composition and defect configuration, assuming that all migration pathways can be derived from this static structure. Variations in migration kinetics due to local chemical environments are captured as distributions of transition statistics. 

A full description of the HopDec method and code will follow; to facilitate this description, we define several key terms:

\begin{itemize}
\item\textbf{State:} An atomic configuration characterized by atom positions, cell dimensions, species type, etc.
\item\textbf{Transition:} An atomic pathway connecting an initial and final \textbf{State}, annotated with kinetic and thermodynamic information.
\item\textbf{Connection:} At least one \textbf{Transition} linking two arbitrary \textbf{States}.
\item\textbf{Decoration:} A specific mapping of user-defined chemical compositions onto a \textbf{Connection}, \textbf{Transition}, or \textbf{State}.
\item\textbf{Redecoration:} The application of multiple \textbf{Decorations} onto \textbf{States}/\textbf{Transitions}, followed by recomputation of pathways and energetics using Nudged Elastic Band (NEB) calculations.
\item\textbf{Model:} The parent object containing all \textbf{Redecorations} and \textbf{Connections} produced during a HopDec execution.
\end{itemize}

A central function of HopDec is to map the defect state-space using a light implementation of methods similar to those described by Swinburne et al. \cite{Swinburne2018} and incorporated in the TAMMBER code \cite{TAMMBER}. Under the assumption of Harmonic Transition State Theory, defect migration trajectories can be coarse-grained into a graph representation where nodes correspond to stable defect minima and edges to transitions with associated kinetic rates. This abstraction enables kinetic modeling, such as Kinetic Monte Carlo (KMC) simulations, to access timescales far beyond those achievable by MD.

In simple materials, defect transport can be mapped using crystal symmetry to identify a limited number of unique \textbf{States} and \textbf{Transitions}. However, in CCMs, symmetry is broken by local compositional disorder, leading to a vastly larger and more varied state-space. HopDec addresses this by coupling state-space mapping with a redecoration campaign, wherein atomic species are randomly assigned to atomic positions according to a user-defined composition. This approach allows HopDec to record variations in defect energetics across randomly sampled local environments, resulting in a transition network where each edge encodes a distribution of rates rather than a single value. While the sampling does not explicitly weight configurations by their thermodynamic probabilities, HopDec records this information, enabling users to incorporate thermodynamic relevance if desired.

\begin{figure*}
    \centering
    \includegraphics[width=\linewidth]{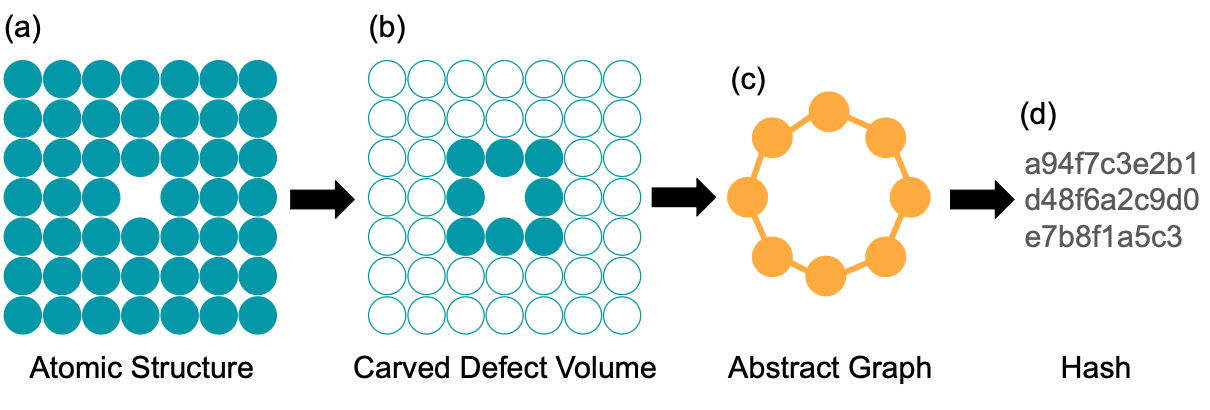}
    \caption{Simplified schematic of carving and defect identification methodology. (a) represents a full atomic structure with a vacancy in some position. (b) Using user-defined centro-symmetry parameters, the defect volume is identified (filled circles) (c) 1NN bonds between atoms are used to generate an abstract graph where a node is an atom and an edge is a bond. (c) An example hash generated from the Weisfeiler-Lehman algorithm, unique to that defect symmetry.}
    \label{fig:Carve}
\end{figure*}

The state-space mapping proceeds as follows: given an initial defect configuration with some fixed \textbf{Decoration}, an interatomic potential, and user-defined parameters, HopDec employs molecular dynamics (MD) through the LAMMPS-Python interface \cite{Thompson2022} to discover new stable defect \textbf{States}. A new \textbf{State} is considered discovered when any atom moves beyond a user-specified displacement cutoff following local energy minimization, akin to detection methods used in accelerated MD techniques such as Temperature Accelerated Dynamics (TAD) \cite{Zamora2016} and Parallel Replica Dynamics (ParRep) \cite{Voter1998}. \textbf{States} are characterized within a defect-centered volume, carved out based on centro-symmetry parameters, and uniquely labeled using a Weisfeiler-Lehman graph hashing algorithm \cite{Shervashidze2011} implemented via NetworkX \cite{Hagberg2008}. A schematic of this process can be seen in Figure \ref{fig:Carve} and ensures that, given sensible centro-symmetry parameters, symmetrically equivalent defects will have the same hash string and can be identified efficiently. A similar strategy is employed to identify symmetrically unique \textbf{Transitions}.

When a new \textbf{State} is identified, a Nudged Elastic Band (NEB) calculation \cite{Henkelman2000} is automatically triggered to compute the energy barrier of the corresponding \textbf{Transition}. Resolving NEB pathways from MD trajectories is non-trivial due to the possibility of multiple concurrent transitions during an observation window. To manage this, HopDec wraps the Atomic Simulation Environment's (ASE) \cite{Larsen2017} implementation of the NEB method with additional layers for automatic intermediate minimum detection, queuing, and path validation, ensuring reliable resolution of transitions.

This iterative procedure progressively builds a defect state-space graph, with a level of completeness dependent on the simulation parameters. The resulting data can be used directly for various analyses, including visualization of migration pathways, evaluation of migration barriers, and coarse-grained kinetic modeling.

Because the initial state-space mapping is based on a single static \textbf{Decoration}, HopDec extends its capabilities with a \textbf{Redecoration} algorithm. For each discovered \textbf{Transition}, a user-defined number of chemical \textbf{Decorations} is generated by randomly assigning species to atomic sites, followed by recalculation of the migration energetics. Each decorated \textbf{Transition} captures the influence of a specific local chemical environment on the migration barrier and defect stability. This process populates the HopDec \textbf{Model} object with an extensive set of decorated \textbf{Transitions} and associated energy distributions. 

The goal is to quantify the mobility of a given type of defect in a complex alloy. Ultimately, this information would be integrated into mesoscale models where the defect-defect interactions, such as recombination and clusters, would be handled. Thus, HopDec provides critical kinetic and thermodynamic information for such models.

HopDec is explicitly designed for massive parallelization on high-performance computing platforms. During state-space mapping, independent MD trajectories are assigned to separate MPI ranks, enabling rapid exploration of defect transitions and configurations. Likewise, the \textbf{Redecoration} phase operates in an embarrassingly parallel manner, where each NEB calculation is independent and communication is only required upon completion when data is gathered to the master rank.

\begin{figure*}
    \centering
    \includegraphics[width=\linewidth]{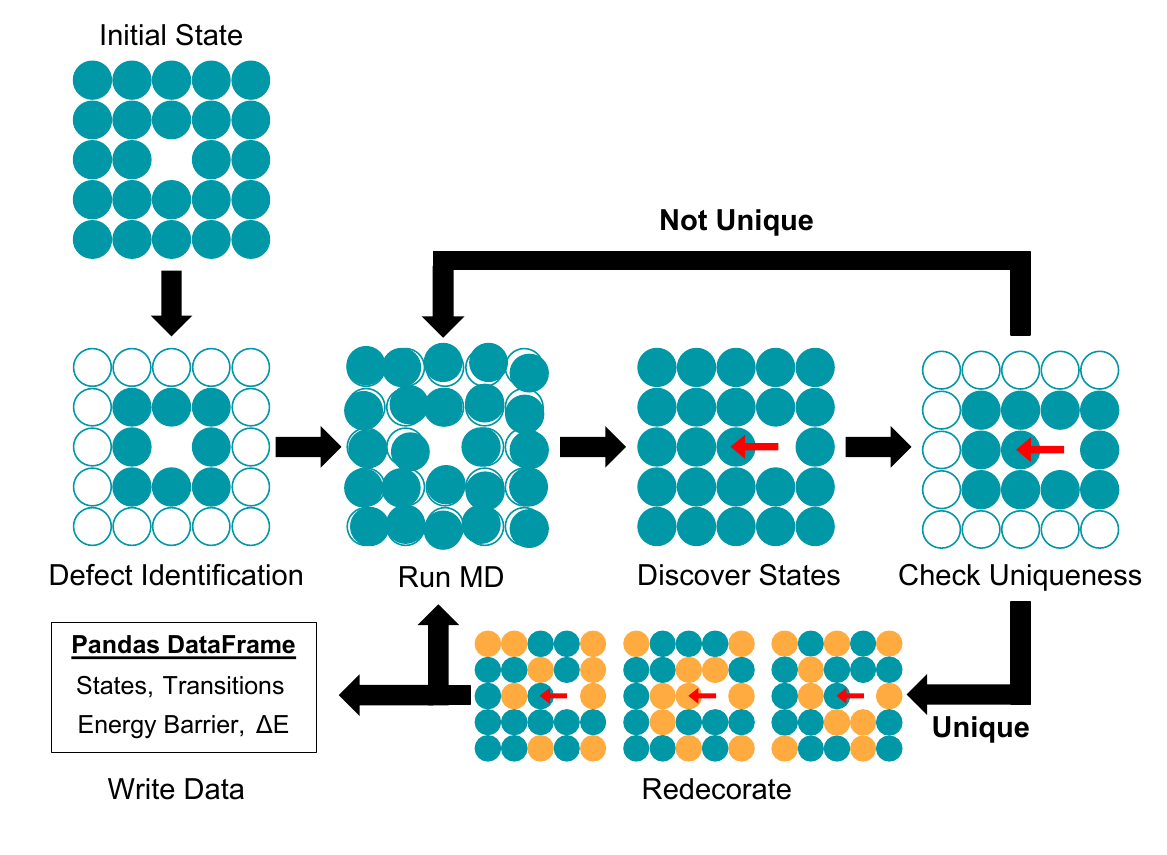}
    \caption{A simplified schematic diagram of the full HopDec operation.}
    \label{fig:HD}
\end{figure*}

A simplified schematic diagram of the full HopDec operation is given in Figure \ref{fig:HD}.

\subsection{Implementation Details}

Under the hood, HopDec uses LAMMPS \cite{Thompson2022} as the MD driver, requiring users to specify LAMMPS-style input parameters. For the case studies presented herein, the Cu-Ni study employed an embedded-atom method (EAM) \cite{Daw1984} interatomic potential for Cu-Ni alloys \cite{Foiles1985}, while the (Fe,Ni)Cr$_2$O$_4$ study utilized a Buckingham–Coulomb potential plus a Morse potential form by Chartier et al. \cite{Chartier2013} previously validated for defect transport simulations in spinel structures \cite{Hatton2024a,Hatton2023,Chartier2008,vanBrutzel2019}. This interatomic potential assigns fixed partial charges to each species: Fe/Ni: 1.2+, Cr: 1.8+, and O: 1.2- reflecting the ionic behavior of the underlying spinel. As a result, defects such as vacancies and antisites are effectively charged. While the model does not account for charge transfer effects, this is not expected to significantly impact the properties studied here. In contrast to systems like Fe$_3$O$_4$ where Fe$^{2+}$/Fe$^{3+}$ redox plays a major role \cite{Srivastava2023}, charge transfer in (Fe,Ni)Cr$_2$O$_4$ is believed to be minimal. Moreover, fixed-charge models have previously captured cation disorder in similar systems \cite{Yamamoto2008,Chartier2008,vanBrutzel2019,Hatton2023}, suggesting that charge transfer has a limited effect on the electrostatic landscape driving disorder \cite{Stevanovic2010}. Additionally, contributions to the total energy from net charges in the system will cancel out identically for $\Delta$E and will not contribute to forces for NEB calculations.

Energy minimizations are performed using the conjugate gradient algorithm with force tolerances converged to machine precision (10$^{-20}$ eV/Å), while NEB calculations used 11 images, a spring constant of 0.1 eV/Å, and a NEB force convergence criterion of 0.01 eV/Å. These parameters are adjustable by the user through the HopDec interface, provided they are compatible with LAMMPS and ASE requirements. Where appropriate, key HopDec configuration settings are indicated in italics.

HopDec is initialized from an atomic structure containing the defect of interest in isolation, so that its kinetics and thermodynamics can be probed precisely as a function of varying local environment. In Cu-Ni, the defect identification was done through the process outlined in Figure \ref{fig:Carve} with centrosymmetry parameters of \textit{centroN=12} (N nearest neighbors) and \textit{centroCutoff=1} (below which atoms are considered to be in bulk). The complexity of the FeNiCr$_2$O$_4$ structure is such that no unique centrosymmetry parameters could reliably identify a lattice defect. Therefore, HopDec was only invoked in Redecoration mode with a pre-defined transition of the vacancy on the A (tetrahedral) sublattice. Many defect identification methodologies exist outside of centrosymmetry such as reference structure based approaches and geometric/tessellation based methods which would likely have greater success with complex structures such as this, however, these are currently not implemented in HopDec.

In cases where the Redecoration algorithm is employed, 1000 decorations (\textit{nDecorations=1000}) is used. This can sometimes lead to slightly more or less data points since if decorations result in a transition decomposing into multiple hops, each of these will be extracted separately, and if NEBs fail due to some instability in the decoration then no data point will be generated. For example, if during minimization any atom moves more than a user-defined cutoff (typically a few Å's) this will be deemed unstable, similarly if no reasonable energy barrier can be detected.

\section{Results \& Discussion}

Two case studies are presented to demonstrate the execution and analysis of HopDec in the case of alloys in a solid solution and a complex oxide with distinct sublattices, requiring specialized treatment.

Complete input parameter sets used in the case studies are available in the HopDec codebase \cite{HD}.

\subsection{Case Study I: Alloy (CuNi)} \label{sec:CuNi}

In this first case study, we focus on the Cu-Ni binary alloy system as a model environment for studying defect migration in chemically disordered FCC metals. The system offers a unique balance of complete solid solubility, similar atomic radii, and a shared FCC crystal structure which provide a controlled framework, while variations in local bonding environments and migration barriers introduce meaningful configurational complexity. This combination makes Cu-Ni an ideal platform for isolating and understanding the effects of chemical disorder on defect behavior.

By applying HopDec to Cu-Ni alloys, we aim to test and demonstrate its ability to resolve migration pathways and defect energetics across diverse local environments. This work not only serves as a validation study, but also provides insight into how local chemistry modulates defect mobility in substitutional alloys—a phenomenon that underpins the performance of structural materials in radiation environments, high-temperature reactors, and additive manufacturing applications \cite{Walck2024,Miracle2017,Zhang2014, Yang2022,Xiao2022}

HopDec was first applied to pure Cu, mapping the defect state-space for vacancies, di-vacancies, and interstitials at 1200 K over 100 ns of MD, initially without redecoration (\textit{modelSearch=1, redecorateTransitions=0, Temperature=1200, maxMDTime=$10^5$}). Transitions between states during MD occur stochastically and can be modeled as a Poisson process, where the probability of observing at least one event in a time interval K for a process with rate $\lambda$ follows $P = 1 - e^{-\lambda K}$. Based on this model, the 100 ns simulation at 1200 K provides approximately 90 \% confidence that all transitions with effective rates corresponding to barriers below 1.34 eV were sampled though this estimate assumes ideal Poisson statistics and applies only to the pure Cu system under these conditions; it may not hold for Ni or Cu-Ni alloys.

For the vacancy, only the standard FCC hop was observed. For the di-vacancy, we enforced in HopDec that the two constituent vacancies remained $<=$2NN apart, within this range the migration energies will be strongly coupled, thus two mechanisms were found: (i) translation of the defect complex and an overall rotation of the di-vacancy direction and (ii) separation of vacancies from 1NN to 2NN, both consistent with previous work \cite{Fotopoulos2023}. 

For interstitials, both standard rotation and translation mechanisms of the split-interstitial were discovered with energy barriers of 0.21 eV and 0.15 eV, respectively. Alongside this, 2 elongated migration paths were discovered that combined rotational and translational motion within a single migration event and energy barrier. One of these elongated mechanisms transported the interstitial over a distance of approximately $\simeq$ 3.6 Å—significantly farther than the $\simeq$ 2.5 Å typical of standard translation hops—yet exhibited a surprisingly low energy barrier of just 0.18 eV. This involved a 1D migration along a $\langle100\rangle$-type direction, effectively replacing at least two standard translation events. The second spanned three dimensions, equivalent to a sequence of a typical translation and rotation within a single event with a barrier of 0.26 eV. The combination of compound translations and low energy cost suggests that these mechanisms could play an important role in interstitial transport in Cu and potentially other FCC metals. Notably, such pathways have not been previously reported in any FCC metal system, highlighting the capability of automated transition detection methods to uncover non-intuitive yet critical migration events—unconstrained by predefined, hand-crafted pathways. However, during redecoration, these elongated migrations decomposed into multiple events indicating that they were being stabilized by the chemical symmetry of their environment, therefore, only unit rotations and translations are discussed further.

\begin{figure*}
    \centering
    \includegraphics[width=\linewidth]{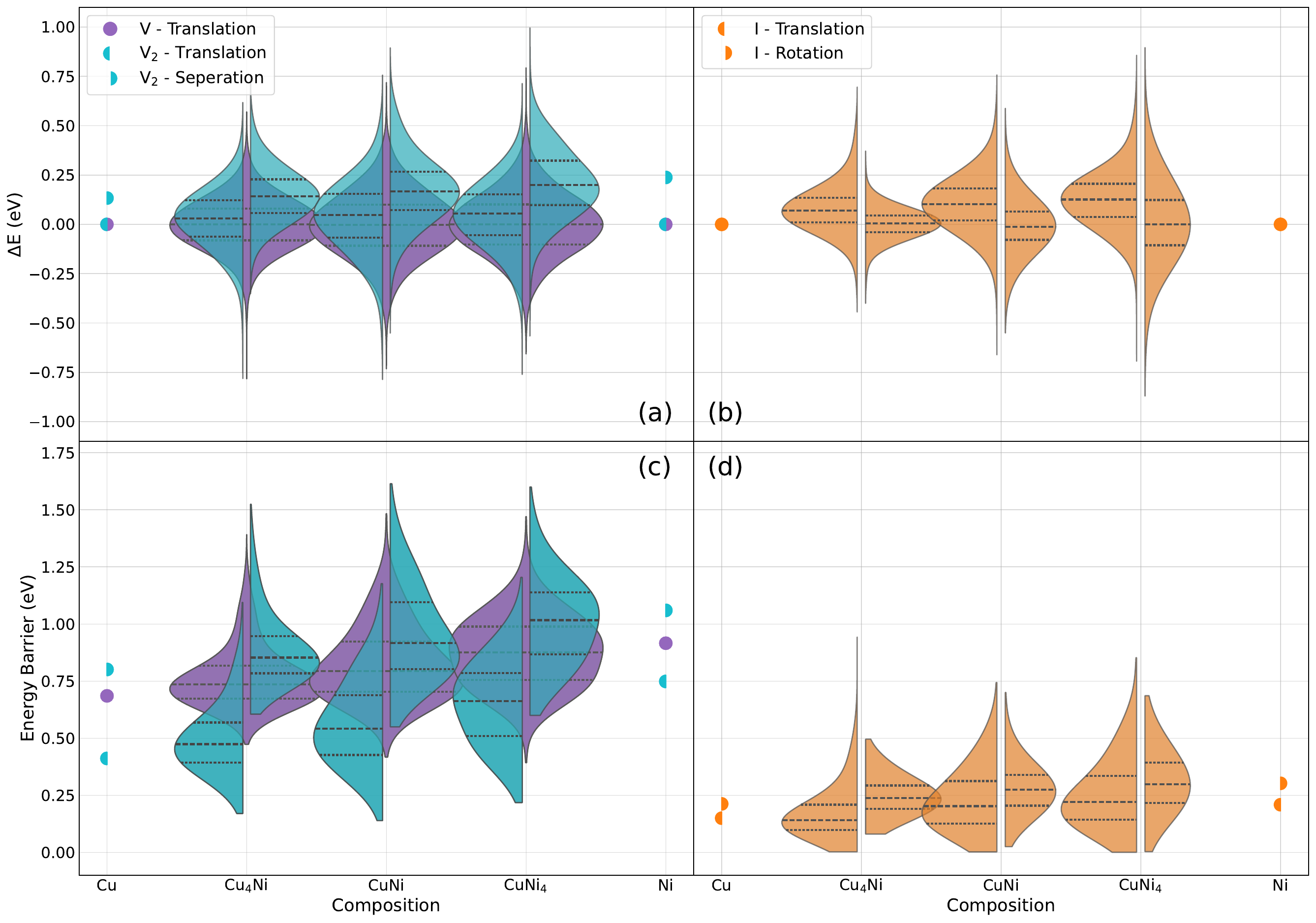}
    \caption{(a,b) Energy change ($\Delta E$) and (c,d) Energy Barrier for migration of defect complexes as a function of Cu-Ni alloy compositions. (a,c) Vacancy complexes and (b,d) Interstitials. Points in the pure metals represent single values for each mechanism and semi-circles correlate to the split violin plots in the intermediate compositions. Violin plots pointing in different directions indicate different mechanisms for the defect and the lines are the mean and standard deviation }
    \label{fig:dE-bar_alldefects_CuNi}
\end{figure*}

Figure \ref{fig:dE-bar_alldefects_CuNi} shows $\Delta$E and E$_\text{b}$ data from the subsequent redecoration campaigns of the identified transitions at varying Cu-Ni compositions (Cu$_{1-x}$Ni$_{x}$). It should immediately be noted how broadly distributed the data is for each case, indicating that the local chemical environment for defects has significant impact on their kinetics and themodynamics. Despite this complexity, an approximate Vegard’s Law trend was observed in the mean migration barrier, suggesting both a simple linear dependence on Ni content and that migration is overall slower with increasing Ni. Additionally, all E$_\text{b}$ distributions became broader indicating that local atomic configurations exert a stronger perturbative effect on defect motion through the creation of a more heterogeneous energy landscape with increasing Ni content. Besides those discussed in the self-interstitial case, no decomposition of single-barrier migration pathways was observed upon redecoration, implying that defects remain effectively on-lattice across the full composition range—a convenient result for potential lattice KMC simulations. If decomposition did occur, HopDec would automatically detect this and recompute NEBs for each sub-transition, returning this information to the user.

For isolated vacancies, the $\Delta E$ distributions remain centered near zero across all compositions, indicating that forward and reverse migration events occur with roughly equal probability. This symmetry suggests that a wide range of local atomic environments has been sampled, with little bias, and that the overall distribution remains consistent across the Cu-Ni composition range.

We analyzed transitions that produced extreme values—defined as those exceeding three standard deviations from the mean—in both $\Delta E$ and $E_b$. Extreme high $E_b$ values were associated with migrating Ni atoms encountering a saddle point where the nearest neighbors (NN) are saturated with Cu atoms. Notably, no transitions exhibited extremely low $E_b$ values. For $\Delta E$, extreme high values typically involved Ni atoms moving from an initial state rich in Ni in the first NN shell but poor in Ni in the second NN shell, to a final state with the inverse arrangement. The opposite trend was observed for extreme low $\Delta E$ values. A more detailed analysis of how local atomic environments influence these energy metrics is provided later in this section.

Di-vacancy $\Delta E$ data also indicates well-sampled local environments and a positive trend in the separation mechanism, indicating that the binding energy becomes increasingly positive with higher Ni content, and therefore enhanced thermodynamic stability of di-vacancy complexes in Ni-rich alloys. Moreover, the migration barriers for bound di-vacancies are significantly lower than for isolated vacancies, suggesting that di-vacancy transport may dominate under certain conditions,  though the rotational di-vacancy mechanism results in a smaller defect displacement than an isolated vacancy. As with isolated vacancies, the analysis of extreme values—those beyond three standard deviations from the mean—reveals similar patterns for both transition types. This consistency supports the idea that local atomic configurations exert comparable influence on both isolated and di-vacancy migration behavior.

For interstitials, the overall picture remains relatively stable throughout the compositional range suggesting that interstitials are the least affected by a changing local chemical environment. Extreme high $E_b$ values, observed across both interstitial migration mechanisms, are characterized by atoms passing through saddle points where the 2NN sites are nearly saturated with Cu. Interestingly, there is no strong correlation with either the migrating species or the composition of the 1NN shell. This is consistent with the nature of split interstitials, which induce larger local displacements than vacancies and therefore may be more influenced by longer-range compositional effects. No consistent pattern in local environment was identified for extreme high or low $\Delta E$ values, further supporting the idea that interstitial thermodynamics may be governed by more extended atomic configurations beyond the immediate neighborhood. Across all compositions, a subset of energy barriers was found to be very close to 0.0 eV. In these cases, HopDec identified small—though sometimes negligible—barriers, suggesting the presence of environments that are only marginally stable. If a site were completely unstable, HopDec would automatically discard it due to relaxation away from the initial defect configuration. However, the fact that these small barriers were recorded implies that the sites are metastable. Supporting this, the corresponding $\Delta E$ values for these cases were all at least two standard deviations below the mean.

Close analysis of this data revealed simple yet powerful relationships between defect energetics and some metrics of local chemical environments. This is exemplified best for isolated vacancies which will now be the focus of the subsequent discussion. In particular, $\Delta$E correlates strongly with two key parameters:  the migrating species (M.S.) (Cu or Ni) and the change in the number of Cu atoms ($N_{\text{Cu}}$) in the 1NN and 2NN shells around the defect site (M1 and M2, respectively):

\begin{equation}
    \Delta N_{\text{Cu}}^{\text{(M1)}} = N_{\text{Cu}}^{\text{(F: M1)}} - N_{\text{Cu}}^{\text{(I: M1)}},
\end{equation}
and similarly for the 2NN shell (M2): 
\begin{equation}
    \Delta N_{\text{Cu}}^{\text{(M2)}} = N_{\text{Cu}}^{\text{(F: M2)}} - N_{\text{Cu}}^{\text{(I: M2)}},
\end{equation}

\begin{figure}
    \centering
    \includegraphics[width=\linewidth]{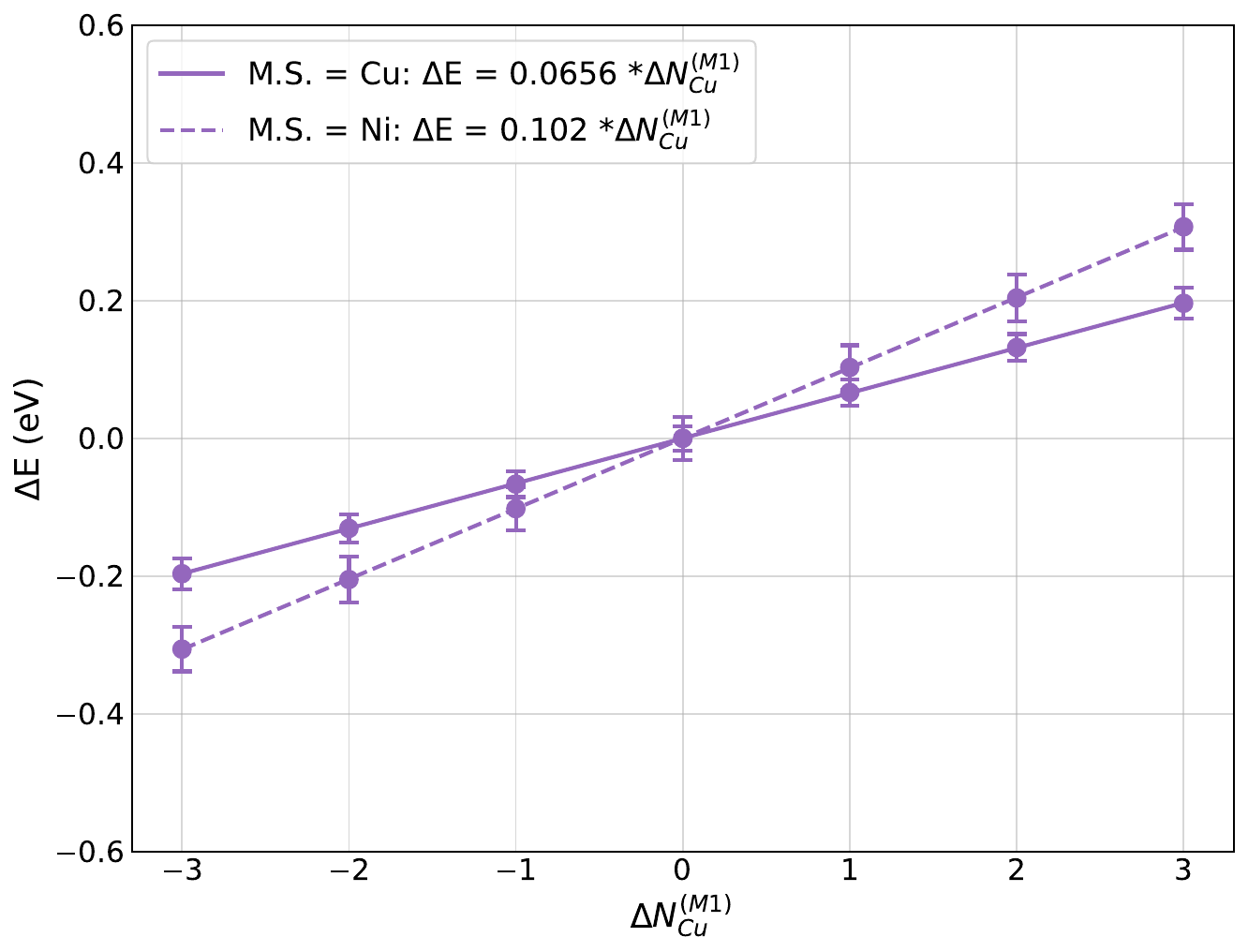}
    \caption{$\Delta E$ values for vacancy migration against the delta Cu count in the 1NN shell at the minima (M1). Circles are mean values and error bars represent 1 standard deviation.}
    \label{fig:dE_V}
\end{figure}

\noindent where I and F refer to the initial and final minimum energy configurations (states), respectively.

This correlation is exemplified in Figure \ref{fig:dE_V} showing $\Delta$E as a function of M.S. and 1NN shell deltas. There is a strong correlation between these parameters and the mean values form a neat linear relationship with a tight standard deviation. While some scatter remains, it is likely attributable to compositional variations beyond the 1NN shell, which have a comparatively smaller influence on thermodynamic stability. Given this, we propose the following equation to approximate $\Delta$E:

\begin{equation}\label{eqn-dE}
\Delta E =
\begin{cases}
\begin{aligned}
& + 0.0627 \cdot \Delta N_{\text{Cu}}^{\text{(M1)}} \\ 
& - 0.00285 \cdot \Delta N_{\text{Cu}}^{\text{(M2)}}, 
\end{aligned} & \text{if M.S. } = \text{Cu} \\
\\
\begin{aligned}
& + 0.0952 \cdot \Delta N_{\text{Cu}}^{\text{(M1)}} \\
& - 0.00774 \cdot \Delta N_{\text{Cu}}^{\text{(M2)}},
\end{aligned} & \text{if M.S. } = \text{Ni}.
\end{cases}
\end{equation}

Testing this fit on the HopDec data resulted in a root mean-squared error (RMSE) of 0.020 eV, which is surprising considering the small number of variables and simplicity of the functional form.

Firstly, these fits imply that a vacancy migrating between states that have identical 1NN and 2NN species counts, have $\Delta$E $\simeq 0$ even when the exact arrangement of the species in those NN shells is different. This strongly suggests, at least from a thermodynamic perspective, that the local environment is almost completely defined by the net chemistry between states. Secondly, according to the fit, regardless of the migrating species, vacancies in Cu-Ni energetically prefer to move to areas of \emph{decreased} Cu concentration in the 1NN positions but \emph{increased} Cu content in the 2NN shell. The differing signs in the dependence on each NN shell is remarkable, indicating that changes in local environment act oppositely depending on their distance from the defect, a highly un-intuitive finding which could potentially be driven by the size variations between the species.

In the case of the migration barrier (E$_\text{b}$), we notice that there is strong dependence on the previously described M.S. and $\Delta N_{\text{Cu}}^{\text{(M1)}}$, as well as on the number of Cu atoms at the 1NN and 2NN positions of the transition saddle point: $N_{\text{Cu}}^{\text{(S1)}}$ and $N_{\text{Cu}}^{\text{(S2)}}$. Fitting a simple linear function to these variables yields the following equation for Energy Barrier (E$_b$):

\begin{equation}\label{eqn-bar}
E_b =
\begin{cases}
\begin{aligned}
& + 0.0318 \cdot \Delta N_{\text{Cu}}^{\text{(M1)}} \\
& + 0.0507 \cdot (N_{\text{Cu}}^{\text{(S1)}} - 4) \\
& - 0.0116 \cdot (N_{\text{Cu}}^{\text{(S2)}} - 22) \\
& + E^{\text{Cu}}_{\text{b}}
\end{aligned} & \text{if M.S.} = \text{Cu} \\
\\
\begin{aligned}
& + 0.0517 \cdot \Delta N_{\text{Cu}}^{\text{(M1)}} \\
& + 0.109 \cdot N_{\text{Cu}}^{\text{(S2)}} \\
& -0.0184 \cdot N_{\text{Cu}}^{\text{(S2)}} \\
& + E^{\text{Ni}}_{\text{b}}
\end{aligned} & \text{if M.S.} = \text{Ni}
\end{cases}
\end{equation}

where $E^{\text{Cu}}_{\text{b}}$ and $E^{\text{Ni}}_{\text{b}}$ are the energy barriers for migration in the pure materials of 0.68 eV and 0.92 eV respectively. Note that the equation has been shifted to recover these barriers in the pure case.  Once more, this simple functional form and a few parameters leads to a reasonably good fit with an RMSE of 0.056, adequate for studying the dominant effects on the energy barrier. As with $\Delta$E, note that the parameter controlling the dependence on Cu near the saddle point is not the same sign for 1NN and 2NN shells, this indicates that the energy barrier depends independently on Cu in each neighbor shell.

These fits for both $\Delta$E and E$_\text{b}$ achieved decent accuracy considering their simplicity, indicating that vacancy energetics in Cu-Ni alloys are governed primarily by simple chemical counts within 1NN and 2NN shells. Most importantly, these fits show that, in general, vacancies migrate to sites with decreased Cu in the 1NN shell but increased Cu in the 2NN shell (equation \ref{eqn-dE}) and the lowest energy barriers are when a vacancy hops towards a site with decreased Cu concentration through a saddle point which is low in Cu in its 1NN shell but high in Cu in its 2NN shell (equation \ref{eqn-bar}). These valuable and highly non-obvious insights would be extremely difficult to guess \textit{a-priori} and could only be available through the automated routines that HopDec provides.

Overall, this case study highlights how HopDec enables rapid, detailed mapping of defect energetics in chemically complex alloys, providing robust data for constructing reduced-order models and informing multiscale simulations. While in this case, a simple linear relation was sufficient to capture the description of local environment, more sophisticated methodologies could be used such as more advanced functional forms and/or the use of machine-learning. Indeed, this type of data has been used previously to construct high-accuracy machine-learned models of vacancy transport in Cu-Ni \cite{Talapatra2024}.

\subsection{Case Study II: Complex Oxide ((FeNi)Cr$_2$O$_4$)}

Spinel-structured complex oxides (A$^{2+}$B$_2^{3+}$O$_4^{2-}$) are of growing interest for advanced energy and nuclear technologies due to their high thermal stability, flexible cation distribution, and the inherent tolerance of some chemistries to radiation-induced defects  \cite{Sickafus1999,Satalkar2016,Kinoshita1995,Robertson1991,Kreller2021}. Spinels consist of two crystallographic cation sublattices: an octahedral sublattice typically occupied by B$^{3+}$ cations, and a tetrahedral sublattice occupied by A$^{2+}$ cations in the so-called normal spinel configuration. The desirable damage tolerance of spinels are primarily a result of the accommodation of point defects through the creation of cation antisites, called spinel inversion \cite{Sickafus1995,Sickafus2000,Uberuaga2015}. The inversion parameter quantifies the extent to which A cations occupy octahedral sites and B cations occupy tetrahedral sites. However, it is important to distinguish inversion from chemical disorder: even a fully inverted spinel can be either ordered or disordered depending on whether cation occupations are systematic or random. Similarly, a chemically complex spinel with a normal cation arrangement may still exhibit sublattice disorder if multiple cation species are randomly distributed on a single sublattice. As damage accumulates (e.g., via irradiation or elevated temperature), the inversion state can evolve, altering cation distributions and enabling non-intuitive migration pathways that significantly affect defect transport \cite{Uberuaga2021,Kreller2021,Uberuaga2007}.

The equimolar Fe$_{0.5}$Ni$_{0.5}$Cr$_{2}$O$_{4}$ (FNCO) system presents an ideal model system for studying the influence of cation chemistry on defect dynamics. Of particular interest is the role of cation vacancy migration in both the normal and random spinel configurations, as cation transport plays a critical role in radiation tolerance, long-term structural stability, and phase evolution. In radiation environments, vacancy-mediated diffusion can significantly impact the material’s resistance to amorphization, its capacity to self-heal, and its ability to accommodate non-stoichiometry.

In this study, we employ HopDec to investigate the mechanisms and energetics of representative defect migration in both normal and random spinel configurations. The migration was hand-made to be a vacancy hop on the tetrahedral sublattice which, in normal spinel, is occupied by A atoms. This is chosen as a representative scenario in which the migration barrier has been shown previously to strongly depend on local chemical environment through inversion \cite{Hatton2023,Hatton2024a}. By quantifying migration barriers and identifying preferred diffusion pathways, we aim to provide insights into how cation ordering influences defect mobility—an essential step toward evaluating this material’s suitability for radiation-tolerant applications. This spinel system exhibits two levels of chemical complexity: first, in the \lq normal' structure, Fe and Ni share the A sublattice, and second, in the random case, Fe, Ni, and Cr occupy both A and B sublattices.

To handle this multi-leveled disorder, HopDec uses the concept of \lq active' and \lq static' species. Active species are those which are to be used in the redecoration algorithm and Static are those that are not-that remain fixed through the redecoration procedure. Therefore, in the \lq normal' spinel case we set \textit{activeSpecies='Fe','Ni'} and \textit{staticSpecies='Cr','O'} so that only Fe and Ni atoms are used in the redecoration algorithm on the A sublattice. In the case of the random Spinel where A and B sites can disorder, we set \textit{activeSpecies='Fe','Ni','Cr'} and \textit{staticSpecies='O'} since we now only keep O fixed on the anion sublattice. These two distinct settings in HopDec allow us to probe two diverse material conditions and understand the impact on defect migration.

\begin{figure*}
    \centering
    \includegraphics[width=\linewidth]{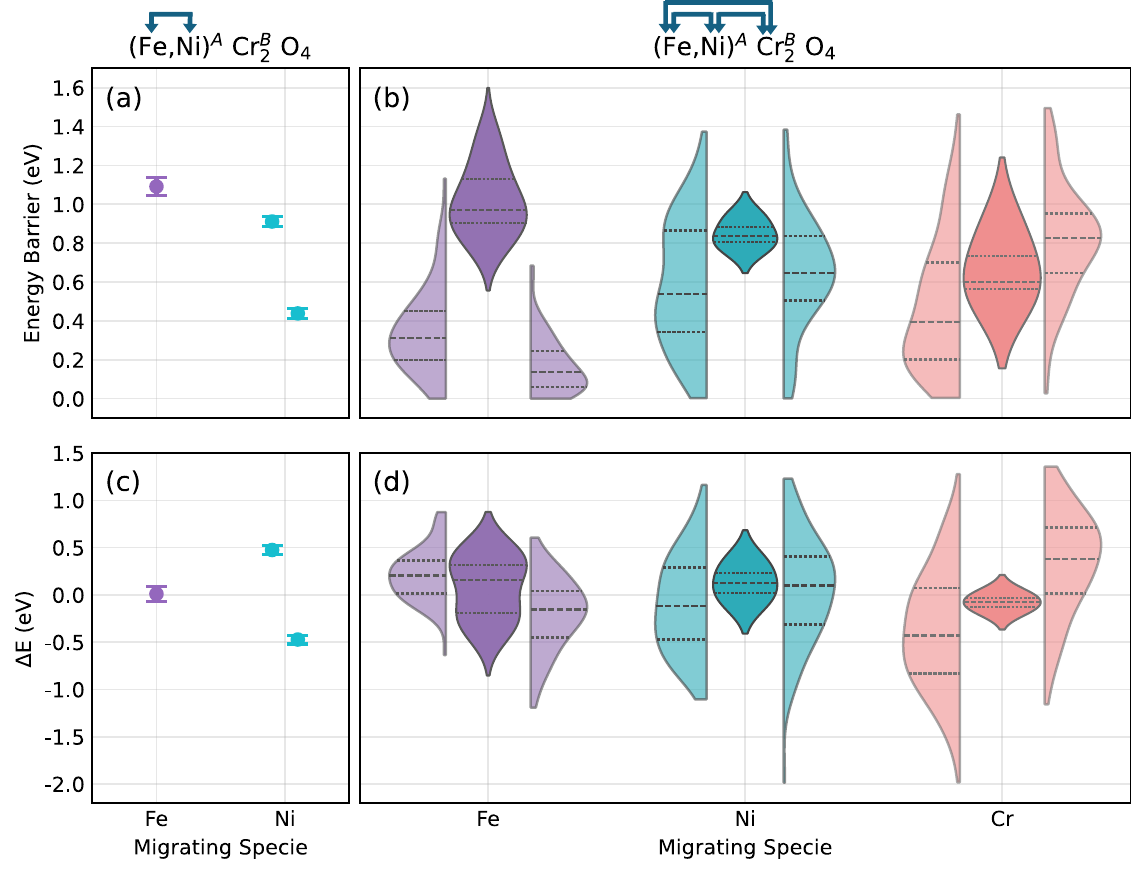}
    \caption{(a,c) Mean and standard deviation of (a) Energy Barrier and (c) $\Delta E$ in normal spinel where disorder arises from changes in Fe and Ni atom distribution. (b,d) Violin plots of (b) Energy Barriers and (d) $\Delta E$ in inverse spinel in which disorder arises from redistribution of atoms across both cation sublattices. Data plotted opaquely on the grid line indicate a single event between initial and final defect state. Data plotted translucently to the right and left of the grid line indicate events that decomposed into two hops between initial and final state. Note that data for the second hop is displayed relative to the intermediate state between the hops.}
    \label{fig:dE-bar_alldefect_FeCrNiO}
\end{figure*}

Figure \ref{fig:dE-bar_alldefect_FeCrNiO} shows $\Delta$E and E$_\text{b}$ for the normal and random spinel configurations as a function of migrating atomic species, found again to strongly influence these properties. Migration events that involve a single, direct saddle point between the initial and final defect configurations are plotted opaquely, centered on the grid line. In contrast, events that proceed via a two-step (compound) hop are plotted translucently, with the first hop shown to the left of the grid line and the second hop to the right. A full migration in these cases is represented by the combination of both hops—the first and second $E_\text{b}$ and $\Delta E$—with the second segments plotted relative to the intermediate state. As a result, a lower barrier for either individual hop does not necessarily imply faster kinetics overall, due to the compound nature of the migration pathway.

In the normal case, there is little variation with tight standard deviations. This is because the 1NN cation shell is that of the B sublattice, which is fixed to Cr in the \lq normal' spinel so only more subtle 2NN affects are seen. Additionally, we observe that if Fe is the migrating species, there is a single transition barrier between the initial and final state of the vacancy, however, when Ni is migrating there are 2, with the intermediate minima representing a split-vacancy which is higher in energy, since for the first transition $\Delta$E is positive and for the second it is negative. This is an interesting finding and points to the variation in defect migration pathways in these complex materials.

In the random spinel, the A \& B sublattices are now occupied by a random share of Fe, Ni and Cr atoms which results in large spreads in the data. Firstly, when Fe is migrating, under some conditions it will decompose into two transitions which have a significantly lower energy barrier. Further, under some local environments, migrating Ni has a single transition, indicating how local chemical environment not only modulates energy barriers but also the stability of migration pathways. Additionally for Ni and Cr, the mean $\Delta$E data suggests that the split vacancy minima is in-fact the more stable configuration for the vacancy, pointing once again to the strong influence of local ordering on defect transport transport mechanisms. As a reminder, for compound hops, the second step is plotted relative to the energy of the intermediate minimum, this is important when interpreting both the total energy change and apparent barrier heights.

In an attempt to understand the nature of these drastic spreads in data for the random spinel, we investigate the local environment of the vacancy, similarly to section \ref{sec:CuNi}. We find no correlation between $\Delta$E and E$_\text{b}$ based on the species counts in the NN shells, in contrast to the results presented in  Section \ref{sec:CuNi}. Indeed, in cases where a vacancy moved between sites with 1NN and 2NN shells which have the same species counts, the $\Delta$E has a large spread of values. This implies that a more sophisticated metric for defining local environment is required. This is quite natural in that, by swapping A and B cations, the local charge homogeneity is thus disrupted through the creation of, effectively, charged antisites which are highly likely to impact defect kinetics/thermodynamics. Indeed, this is also consistent with recent work on similar spinel chemistries which found that the presence of short range order in the disordered inverse structure had a strong impact on defect thermodynamics and kinetics implying that in this case it is more a function of how species are arranged around the defect than simply how many of them there are \cite{Hatton2023}.

While a simple relationship could not be extracted, HopDec still provides both valuable insight and the data structures required to facilitate the construction of more sophisticated models which capture a more complex picture of local environment. Further, raw analysis of this data has revealed how the local chemical environment works to first modify the kinetics and thermodynamics of defect migration and then shift the exact mechanisms which are stable in these complex materials, even changing the stability of the exact defect structure. 

\subsection{Discussion}

The application of HopDec to CCMs across two diverse systems—Cu-Ni alloys, in which chemical complexity arose from the varying alloying concentrations of each metal, and (Fe,Ni)Cr$_2$O$_4$ spinel, where chemical complexity arose both from the multi-element A$^{2+}$ sublattice and the ability of spinels to invert under extreme conditions to give rise to disoder on both cation sublattices. These examples demonstrate the flexibility, robustness, and scientific value of the methodology. In both cases, HopDec effectively explored the vast configurational landscape that would otherwise be inaccessible through conventional atomistic approaches.

In the Cu-Ni alloy, HopDec was able to characterize the variability of vacancy, di-vacancy, and interstitial defect migration energies as a function of local chemical environment. A striking outcome is the identification of simple, predictive relationships between local composition (specifically, the number of Cu atoms in the first nearest neighbor shell) and both the energy difference ($\Delta$E) and the migration energy barrier. These correlations enable reduced-order modeling of defect kinetics, drastically simplifying the parameterization of Kinetic Monte Carlo (KMC) models without sacrificing physical fidelity. Such simplifications are invaluable for multiscale modeling, providing a pathway from atomic-scale simulations to mesoscale predictions of material behavior under operational conditions.

The (Fe,Ni)Cr$_2$O$_4$ spinel case study highlights both the capabilities and the challenges of applying HopDec to more structurally complex oxides. In the normal spinel configuration, migration barriers showed limited spread, consistent with minimal 1NN chemical variability around migrating defects. In contrast, the random spinel exhibited significant variability in migration energies and complex defect pathways, including the formation of intermediate split-vacancy states. Importantly, attempts to correlate migration energetics with simple metrics (e.g., 1NN/2NN species counts) were unsuccessful in the inverse spinel, indicting the need for more sophisticated descriptors, such as local order parameters or graph-based environment representations.

While these case studies do not represent the full compositional complexity typically associated with chemically complex alloys (CCAs) or high entropy systems, the HopDec framework is inherently general. Its design places no restrictions on the number of atomic species, types of sublattices, or crystal structure, and can be readily applied to multi-principal element systems. In fact, the rapid combinatorial growth of distinct local environments with increasing chemical complexity only reinforces the need for automated, scalable workflows like HopDec. These simpler systems were selected to clearly demonstrate and validate the methodology, but future work will focus on more chemically complex systems to fully showcase HopDec’s capabilities in the context of true CCAs.

Moreover, the results reveal future development directions for HopDec. In particular, materials such as UO$_2$, where coupled defect–interstitial motion (e.g., oxygen sublattice distortion during uranium vacancy hops) is important, will require additional methodological enhancements. Extending HopDec to properly capture these correlated atomic movements is essential for its application to fluorite structures and other materials with mobile sublattices. Additionally, HopDec is naturally suited to studying the role of dopants on defect kinetics, a crucial factor in engineered materials where dopant-defect interactions strongly govern macroscopic properties.

Overall, HopDec’s modular and massively parallel design enables efficient exploration of defect landscapes and paves the way for coupling with machine learning approaches. Its ability to generate environment-sensitive data is ideal for constructing accurate reduced-order models (ROM), which are key to bridging atomistic simulations with mesoscale and continuum-scale predictive frameworks. Indeed, a natural extension of HopDec is its application to higher-fidelity, computationally expensive methods such as Density Functional Theory or machine-learned interatomic potentials. In these regimes, where efficient sampling is crucial, HopDec facilitates the extraction of high-quality data for ROM construction which ultimately supports robust multiscale modelling by bridging first-principles calculations with long-timescale predictions.

\section{Conclusion}

In this work, we introduced HopDec, an automated, high-throughput atomistic workflow designed to systematically explore defect migration pathways in chemically complex materials. By combining accelerated molecular dynamics techniques with a novel redecoration strategy, HopDec efficiently generates representative and comprehensive datasets of migration barriers and energy differences across diverse chemical environments.

The application of HopDec to Cu-Ni alloys and (Fe,Ni)Cr$_2$O$_4$ spinels demonstrated its power and versatility. In Cu-Ni alloys, simple, physically interpretable relationships between local composition and defect energetics were uncovered, enabling reduced-order models for kinetic simulations. In complex oxides, HopDec revealed rich and intricate defect migration behavior, underscoring the challenges of modeling systems with high configurational disorder and indicating the need for machine-learning-assisted descriptors.

HopDec represents a significant advancement in atomistic simulation workflows for CCMs, bridging the gap between atomic-scale calculations and mesoscale predictive modeling. It provides a blueprint for the integration of defect data into higher-scale simulations, supports the development of data-driven models, and offers a path toward rational design of materials with tailored defect transport properties. Future work will focus on expanding HopDec’s capabilities to handle complex defect clusters and more sophisticated local environment descriptions, as well as coupling HopDec outputs with machine learning frameworks to further accelerate materials discovery and design. \\

\section*{Code Availability}

The Hop-Decorate code is available as an open-source Python package at https://github.com/PeteHatton/Hop-Decorate under the BSD-3 license. The repository includes example scripts and configuration files used in the case studies presented in this work. Users can reproduce all major results, extend the workflow for new materials systems, and integrate HopDec into high-throughput simulation environments. Any future updates or enhancements to the codebase will also be maintained through this repository.

\begin{acknowledgements} 
PH and DP thank the Institute of Pure and Applied Mathematics (IPAM) at the University of California in Los Angeles (UCLA) for providing a venue for fruitful discussions on these topics during the New Mathematics for the Exascale: Applications to Materials Science long program. This work was supported by the Laboratory Directed Research and Development program of Los Alamos National Laboratory under project number 20220063DR. Los Alamos National Laboratory is operated by Triad National Security, LLC, for the National Nuclear Security Administration of the U.S. Department of Energy (Contract No. 89233218CNA000001).
\end{acknowledgements}

\bibliography{bibliography}{}

\end{document}